\documentclass[]{epl}

\title{Self Trapping of a Single Bacterium in its Own Chemoattractant}
\shorttitle{Self Trapping of a Single Bacterium}
\author{Yoav Tsori \and Pierre-Gilles de Gennes}
\shortauthor{Y. Tsori \and P.-G. de Gennes} 
\institute{Physique de
la Mati\`{e}re Condens\`{e}e\\Coll\`{e}ge de France, Paris,
France}


\begin{document}

\maketitle

\begin{abstract}

Bacteria (e.g. {\it E. Coli}) are very sensitive to certain
chemoattractants (e.g. asparate) which they themselves produce.
This leads to chemical instabilities in a uniform population. We
discuss here the different case of a single bacterium, following
the general scheme of Brenner, Levitov and Budrene. We show that
in one and two dimensions (in a capillary or in a thin film) the
bacterium can become self-trapped in its cloud of attractant. This
should occur if a certain coupling constant $g$ is larger than
unity. We then estimate the reduced diffusion $D_{\rm eff}$ of the
bacterium in the strong coupling limit, and find $D_{\rm eff}\sim
g^{-1}$.

\end{abstract}

\section{Introduction}

Budrene and Berg \cite{BB} studied an initially homogeneous
population of {\it Eshrechia Coli} ({\it E. Coli}) bacteria on an
agar plate, in conditions where food (succinate) is available.
Depending on the food content they discovered various patterns
such as moving rings or aggregates. These patterns were lucidly
interpreted by Brenner, Levitov and Budrene \cite{brenner}. They
observed that in the (usual) conditions of rapid diffusion, the
bacteria produce a concentration field of the chemoattractant,
$c(r)$, which has the form of a gravitational field ($c(r)\sim
1/r$ in three dimensions). The bacteria attract each other and
cluster by a ``gravitational'' instability. However, when a
cluster (``star'') is formed, food is depleted and the star then
becomes dark in the center, a ring is created, etc.

Brenner {\it et. al.} also discussed the spontaneous aggregation
of a small group of bacteria, and found that they shall indeed
aggregate if their number $N$ is larger than a certain limit
$N^*$. Because of the (rough) similarity with astrophysics, they
called $N^*$ the Chandrasekhar limit.

Our aim here is to discuss some properties of these small clusters
and in particular the limit of a {\it single} bacterium. We point
out that it may be trapped in its own cloud of asparate. This
question has some weak similarity with a classical problem of
solid-state physics and field theory, the {\it polaron} problem,
defined first by Frolich \cite{frolich} and analyzed by many
theorists \cite{LP,pekar,feynman}. A polaron is an electron
coupled to a phonon field in a solid. In the strong coupling limit
analyzed by Pekar \cite{pekar}, the electron builds up a distorted
region, and is essentially occupying the lower bound state in the
resulting (self-consistent) potential. However, the electron moves
slowly: it has a large effective mass.

Our problem here is somewhat similar: if a certain coupling
constant $g$ is larger than unity, the bacterium sees a strong
attractant cloud. We shall see that in three dimensions, it can
always escape, but in one and two dimensions it cannot. The
question of interest is then the {\it effective} diffusivity of
the bacterium.

In section II we start from the basic coupled equations for the
bacteria and attractant \cite{brenner,KS} (except for an
alteration of the food kinetics). From this we investigate the
possibility of a self-trapped state, define a coupling constant
$g$ and find that it can indeed be of order unity in some
favorable cases. $g$ is (except for coefficients) equal to
$1/N^*$, where $N^*$ is the Chandrasekhar limit of ref.
\cite{brenner}. We are interested in high $g$ values ($N^*<1$). If
there are a number $N$ of bacteria in one small droplet, $g$ is
multiplied by $N$ and the large $g$ limit becomes easier to reach.
In section III we discuss this high $g$ limit and the renormalized
diffusion constant.

\section{Self-trapping}

{\it E. Coli} colonies enjoying a large supply of food are
governed by two coupled reaction-diffusion equations
\begin{eqnarray}
\frac{\partial\rho}{\partial t}&=&-\nabla{\bf J}\hspace{2cm}
{\bf J}=-D_b\nabla\rho+\kappa\rho\nabla c\\
\frac{\partial c}{\partial t}&=&D_c\nabla^2c+\beta\rho\label{c_t}
\end{eqnarray}
$\rho$ and $c$ are the number densities of the bacteria and
chemoattractant fields, respectively, $D_b$ and $D_c$ are the
diffusion constants of the bacteria and attractant, $\kappa$
determines the strength of positive feedback and $\beta$ is the
production rate of the attractant by the bacteria.

Below we are interested in the case of fast attractant diffusion.
In this limit, Eq. (\ref{c_t}) shows us that $\rho$ and $c$ relate
to each other like charge density and potential in electrostatics.
In this analogy $\nabla c$ is the force acting on a bacterium. For
one bacterium at the origin and in two dimensions we find that
\begin{eqnarray}
c=c_0\ln\left(\frac{r_0}{r}\right)+const.
\end{eqnarray}
Here $r_0$ is the cutoff length. It is instructive to consider the
steady-state obtained when ${\bf J}=0$. We find that $\rho$ obeys
the ``Boltzmann distribution''
\begin{eqnarray}
\rho=\tilde{\rho}_0\exp\left(\frac{\kappa
c}{D_b}\right)&=&\rho_0\left(\frac{r_0}{r}\right)^g\\
g&=&\frac{\kappa c_0}{D_b}
\end{eqnarray}
Therefore, a self-trapped state exists if the coupling constant is
$g\geq 2$. In order to know the value of $g$, we denote $e$ the
thickness of the growth medium on the Petri dish, and equate the
attractant flow out of a circular domain of radius $r$ with the
attractant production,
\begin{eqnarray}
eD_c\cdot 2\pi r\nabla c\simeq 2\pi eD_c c_0=\beta
\end{eqnarray}
Thus, the coupling constant $g$ can be written as
\begin{eqnarray}
g=\frac{\kappa\beta}{2\pi e D_b D_c}
\end{eqnarray}
Putting reasonable values for the parameters \cite{brenner}
$D_b\simeq 6.6\cdot 10^{-6}$ cm$^2$/s, $D_c=D_b$, $\beta=10^3$
molecules/bacterium/s, $\kappa\simeq 10^{-14}$ cm$^5$/s and
$e=0.05$ cm, we find that $g\approx 0.73$.

This estimate shows that the coupling between the bacterium and
its own chemoattractant field can be rather strong in many
experimental situations. Calculation along similar lines for the
one-dimensional infinitely long ``wire'' with diameter $d$ gives
$c\sim |x|$ and confined bacterium,
$\rho=\rho_0\exp(-2\beta\kappa|x|/\pi d^2 D_bD_c)$. In three
dimensions, however, $c\sim 1/r$ and a self-trapped state does not
exist because $\rho\sim \exp(\kappa c/D_b)$ does not tend to zero
at large distances. As we will see in the next section, the
coupling of a moving bacterium with its chemoattractant leads to
the appearance of a ``drag force'' acting on the bacterium, which
is manifested by an effective, smaller, diffusion constant.

\section{Effective mobility of a self trapped bacterium in a film}

We have seen above that in favorable situations the coupling
constant can be large, $g\gg 1$. This case occurs, for example,
with a small drop (with diameter comparable to the thickness of
the culture medium) containing a significant number of bacteria.
In the following we consider the bacteria moving at a constant
small velocity $v$, and look for the effective diffusion
coefficient.
The attractant profile is given by
\begin{eqnarray}
\frac{\partial c}{\partial t}=D_c\nabla
^2c+\frac{\beta}{e}\delta(x-vt)
\end{eqnarray}
We write $c(r)$ as $c=\int c_{\bf k}\exp(i{\bf k}({\bf
r}-vt))~{\rm d}{\bf k}$ and obtain for the Fourier component
$c_{\bf k}$
\begin{eqnarray}
c_{\bf k}=\frac{\beta}{D_ck^2-i{\bf kv}}\simeq
\frac{\beta}{eD_ck^2}\left(1+\frac{i{\bf kv}}{D_ck^2}\right)
\end{eqnarray}
The ``force'' $f$ acting on the bacteria is
%
\begin{eqnarray}
f=\nabla c=\int i{\bf k}c_{\bf k}~{\rm d}{\bf k}=\frac{\beta
}{eD_c^2}\int \frac{{\bf k}({\bf k v})}{k^4}~{\rm d}{\bf
k}=\frac{\pi\beta v}{2eD_c^2}\ln(k_{\rm max}/k_{\rm min})
\end{eqnarray}
Approximating the logarithmic term by unity, we identify the
effective mobility $\kappa_{\rm eff}$ as
\begin{eqnarray}
\kappa_{\rm eff}=\frac{v}{f}=\frac{2eD_c^2}{\pi\beta}
\end{eqnarray}
We may return for a moment to a problem of many bacteria with
concentration $\rho({\bf r})$ moving in an external concentration
field $c_{\rm ext}({\bf r})$. The bacteria density at steady state
$\rho=\rho_0 \exp(\kappa c_{\rm ext}/D_b)$ is the same if it is
written in terms of the ``effective'' quantities $\kappa_{\rm
eff}$ and $D_{\rm eff}$ instead of $\kappa$ and $D_b$. This means
that
\begin{eqnarray}
\frac{\kappa_{\rm eff}}{D_{\rm eff}}=\frac{\kappa}{D_b}
\end{eqnarray}
This relation tells us that the effective diffusion constant
$D_{\rm eff}$ is given by
\begin{eqnarray}
D_{\rm eff}=\frac{D_c}{\pi^2 g}
\end{eqnarray}
Hence, bacterial diffusion is greatly diminished because of the
chemoattractant cloud which is left behind.

\section{Discussion}

Isolated bacteria moving in thin films (e.g. in the dental plaque)
may be slowed down by their own chemoattractant at scales larger
than the film thickness. This should be observable in experiments
using fluorescent bacteria. In addition, clusters of a few
bacteria are nearly stopped; this could be relevant for their
ultimate fixation in the plaque.

We discussed some effects of the chemoattractant cloud. One may
wonder whether there is an analogous effect related to the food
problem: the bacterium eats some food, and this creates a depleted
food region around it. If this ``food hole'' is lagging behind the
bacterium, there will be more food available ahead, and the
bacterium can go faster. (The corresponding transport is
reminiscent of a hot wire anemometer). However, the resulting food
effect goes like $v^2$ (not $v$) and is thus irrelevant for our
problem of mobility at low $v$.

\acknowledgments We would like to thank P. Silberzan for
introducing us to the chemoattractant problems and also E.
Rapha\"{e}l for useful comments and discussions.

\end{document}